\documentclass[letterpaper, 10 pt, conference]{ieeeconf}  

\IEEEoverridecommandlockouts                              

\usepackage{url}
\usepackage{graphicx}

\usepackage{epstopdf}
\usepackage[font=footnotesize]{caption}
\usepackage[font=footnotesize]{subcaption}

\usepackage{color}
\usepackage{transparent}

\usepackage{multirow}

\usepackage{array}
\newcolumntype{L}[1]{>{\raggedright\let\newline\\\arraybackslash\hspace{0pt}}m{#1}}
\newcolumntype{C}[1]{>{\centering\let\newline\\\arraybackslash\hspace{0pt}}m{#1}}
\newcolumntype{R}[1]{>{\raggedleft\let\newline\\\arraybackslash\hspace{0pt}}m{#1}}
\usepackage{morefloats}

\title{\LARGE \bf
Health Figures: \\ An Open Source JavaScript Library for Health Data Visualization*.
}

\author{Andres Ledesma$^{1}$, Hannu Nieminen$^{1}$, Mohammed Al-Musawi$^{1}$
    \thanks{*This research was supported jointly by TEKES (the Finnish Funding Agency for Technology and Innovation) as part of the Digital Health Revolution and FiDiPro (Finland Distinguished Professor Programme) projects and the European Commission and TEKES under the ARTEMIS-JU WithMe project.}
    \thanks{$^{1}$Andres Ledesma, Hannu Nieminen and Mohammed Al-Musawi are with the Personal Health Informatics research group as part of the Department of Signal Processing, Tampere University of Technology, Tampere, Finland}%
  		\thanks{{\tt\small andres.ledesma@tut.fi,}}
  		\thanks{{\tt\small hannu.o.nieminen@tut.fi,}} 
  		\thanks{{\tt\small mohammed.al-musawi@student.tut.fi}}%
}

\begin{document}

 \maketitle
\thispagestyle{empty}
\pagestyle{empty}

\begin{abstract}

The way we look at data has a great impact on how we can understand it, particularly when the data is related to health and wellness. Due to the increased use of self-tracking devices and the ongoing shift towards preventive medicine, better understanding of our health data is an important part of improving the general welfare of the citizens. Electronic Health Records, self-tracking devices and mobile applications provide a rich variety of data but it often becomes difficult to understand. We implemented the hFigures library inspired on the hGraph visualization with additional improvements. The purpose of the library is to provide a visual representation of the evolution of health measurements in a complete and useful manner. \\

We researched the usefulness and usability of the library by building an application for health data visualization in a health coaching program. We performed a user evaluation with Heuristic Evaluation, Controlled User Testing and Usability Questionnaires. In the Heuristics Evaluation the average response was 6.3 out of 7 points and the Cognitive Walkthrough done by usability experts indicated no design or mismatch errors. In the CSUQ usability test the system obtained an average score of 6.13 out of 7, and in the ASQ usability test the overall satisfaction score was 6.64 out of 7. \\

We developed hFigures, an open source library for visualizing a complete, accurate and normalized graphical representation of health data. The idea is based on the concept of the hGraph but it provides additional key features, including a comparison of multiple health measurements over time. We conducted a usability evaluation of the library as a key component of an application for health and wellness monitoring. The results indicate that the data visualization library was helpful in assisting users in understanding health data and its evolution over time.

\end{abstract}

\section{Introduction}
The ongoing shift from reactive to preventive medicine requires that citizens have the skills and means to take an active role in developing and maintaining their wellness. Use of self-tracking devices and personal wellness applications is more and more common, ranging from sports tracking applications to personal genome sequence analysis services. These services and devices produce large amounts of data. In addition, Electronic Health Records are increasingly replacing paper records in hospitals and clinics around the world. The combination of these large and heterogeneous data sources is expected to provide a ``predictive, preventive, personalized and participatory'' ecosystem for the benefit of the general welfare \cite{Hood2012613}.

To better understand our health, we need to combine heterogeneous data sources and present the information to the user in a complete and accurate manner. In order to accomplish this, health information technologies and visualization design need to be integrated \cite{lesselroth2011data}. The goal is to provide tools for individuals to take better  decisions regarding their health. Similarly, doctors and other medical experts need tools and solutions for getting a complete and accurate view of the patients health, combining together patient's own measurements and clinical data.

An innovative approach for health data visualization is the Health Graph (hGraph), released publicly by MITRE corporation \cite{follett2012hgraph}. In an earlier study \cite{EMBCLedNiem}, it was found that the hGraph-type radial plot can enhance deep understanding of health data and enable the user to create meaningful health insights based on the interrelationships between the measurements. However, the hGraph shows a static overview of a persons wellness. Disease and wellness are processes that change over time. It is also essentially important to be able to follow up the trajectories in the different parameters, what is the rate of the change and how they respond to events such as medical care actions and interventions. Thus, in addition to a static snapshot such as in hGraph, a temporal way of presenting the data is also needed. While hGraph visualization is useful for the purpose of understanding with a quick glance the overall situation, it lacks features such as a clear division of measurements according to their category, a distribution of the labels to avoid clutter and the notion of time or \textit{evolution} of the data \cite{EMBCLedNiem}. 


 
In this article we present a visualization library based on extending the core ideas of the hGraph. Aim of the library is to provide tools to assist experts and non-experts in the decision making process of assessing the situation of a patient and its evolution over time. To test the user satisfaction, ease-of-use and usefulness of the solution, we built an application for health monitoring using the visualization library. The application is a visualization tool that shows a health coaching program and its effects on the evolution of health and wellness of a modeled patient. We evaluated the software solution using usability tests (Heuristic Evaluation, Controlled User Testing and Usability Questionnaires).  


First in this article we review the state-of-the-art on health data visualizations and describe the hGraph and results related to its usability and usefulness, which motivated us to develop hGraph further. We named the new library built on top of the hGraph core ideas as hFigures. The article details the design and implementation of the hFigures library, the features we implemented and how they address the users' needs. We also describe the usability test process we utilized to assess the library in the context of an application for health monitoring. We present the results and discuss the further improvements of the library and its pitfalls. We also discuss how this library can be used in Personal Health Informatics and in the Health Care processes.

\section{Background}



Visualization tools have mostly focused on Healthcare Information Systems and Electronic Health Records (EHR) \cite{lesselroth2011data}. For instance, TimeLine is a software developed to retrieve data from several sources and presented in a hierarchical and timeline based structure where clinicians can browse chronologically through existing EHRs including MRI \cite{bui2007information}.

Additionally, the growing market for mobile health applications (mHealth) have drawn the attention of researchers, developers and investors \cite{mHealth2014}. These applications provide large volumes of personal health data. While the market and demand are expected to grow, the use of the data has the potential to contribute to a better understanding of our health.

Goetze \cite{goetz2015} demonstrated the impact of data visualization as means to represent health data in a complete and accurate manner. He conducted a project that redesigned laboratory test results from numerical tables into colored graphics. He demonstrated that the patients were able to understand better their health situation when presented with the new designs.

Data integration for health monitoring as a Big Data process for personalized medicine has been approached by Idris \textit{et al}. \cite{7072838}. The visualization of this information uses traditional bar and pie charts to report to the user a historical view of a variety of data including mental, social, physical aspects. The novelty of this work is the integration of heterogeneous data sources while the presentation of the information was done following existing graphical representations.

An extensive choice of graphical representation is listed and explained by S. Few \cite{few2006information}. These techniques have been studied and used widely among researchers and individuals alike. Examples include: bar, stacked bar, line and bullet graphs. These visualizations can be combined to provide a personalized wellness indicator system, as proposed by Soomlek and Benedicenti \cite{Soomlek:2013:AWI:2532685.2532686}.

DeRidder \textit{et al.} developed a combined approach that retrieves data from Personal Health Records (PHRs), and presents them to individual patients using a ``3D medical graphical avatar'' \cite{de2013web}. The solution is built using HTML5 and WebGL to render 3D graphics using the web browser. Patients can browse ``regions of interests'' on their avatar and explore further the information contained in EHRs as well as in PHRs.

As stated by Shneiderman \textit{et al.} \cite{shneiderman2013improving}, new visual representations are needed for ``systematic yet flexible visual analytics processes''. We present an existing tentative solution known as the hGraph, released publicly by MITRE corporation. We describe its main advantages and how they can address these challenges. In a previous study \cite{EMBCLedNiem} we identified possible improvements and based on our own implementation we addressed these issues and extended the features of the hGraph to better address large data sources.

However, a data visualization is only as good as the ability of the intended audience to decode graphical objects into numerical values which conveys a clear message. Therefore, we have to consider the graphical perception of the users who will benefit from the data visualization. Graphical perception is the ability of an individual to decode the information displayed as graphical objects \cite{baird1978fundamentals}, it has been a widely researched field \cite{cleveland1984graphical}. Graphical perception affect how we understand visualized information. In the context of health data, it remains a challenge to design graphical representations for non-medical experts. Graphical representations in this context should enhance the users' ability to understand their health situation and take informed decisions. With this ``deep'' understanding on the health situation, individuals can move towards healthier behaviors.

The graphical representation of health data requires a complete and accurate overview of a patient often including large amounts of measurements, which in turn translates to large datasets. Therefore, health data visualizations need to scale to accommodate large datasets. 

\begin{figure*}[tbh]
\centering
\includegraphics[scale=1]{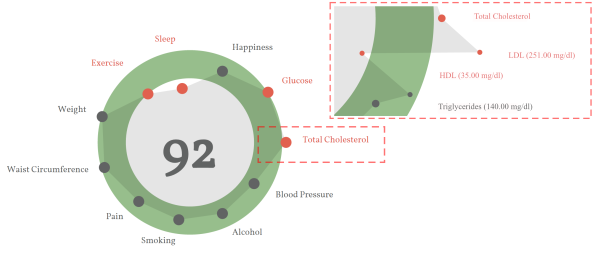}
\caption{An example of the hGraph. The hGraph shows an overview at a low zoom level. When the zoom increases, the details of each measurement are revealed.}
\label{figureHGraphZoom}
\end{figure*}

\subsection{Radar Visualization}


Radar visualization scales to large amounts of data entries because the points are distributed among the circumference. Bar charts, lines or scattered plots, pie charts and similar visualizations can quickly grow to a large scale with large datasets. In the case of radar visualizations, the graphical representation scales with the number of points plotted causing the circumference to grow in order to accommodate all the plotted values. 


Hoffman \textit{et al}. seem to have coined the term of ``radial visualization'' \cite{hoffman1997dna, draper2009survey}. The term was the foundation of `` pie chart, starplot, and radar plot'' which are the basis of ``virtually all the radial visualization methods found in the state-of-the-art research'' \cite{draper2009survey}.

Draper \textit{et al}. propose a classification of radial visualizations into ``three main divisions'' each comprised of ``design patterns'' \cite{draper2009survey}.

\paragraph{Polar Plots}
These are radar visualizations where the center is the starting point from which ``line segments'' originate \cite{draper2009survey}. According to this classification, polar plots are divided into \textbf{Tree} and \textbf{Star} patterns.

Trees have segments that ``branch off'' and are mostly used for visualizing hierarchical data \cite{draper2009survey}. Examples include \textit{Moiregraph} \cite{draper2009survey, jankun2003moiregraphs} and the \textit{Hyperbolic Browser} \cite{draper2009survey, lamping1996hyperbolic}.

Stars do not have branches but rather straight segments originating from the center, their common uses include ``ranking of search results'' and ``viewing relationships among disparate entities'' \cite{draper2009survey}. Examples of star patterns include  \textit{Starstruck} \cite{draper2009survey, hetzler1998multi} and \textit{Neighbourhood Explorer} \cite{draper2009survey, Spence2001Information}.

\paragraph{Space Filling}
Also referred to as \textit{Radial Space Filling} (RSF) \cite{draper2009survey, yang2002interring}, this category comprises the visualizations that fill ``most or all of the area of a circle'' \cite{draper2009survey}. The classification identifies three patterns: \textbf{Concentric}, \textbf{Spiral} and \textbf{Euler}. These patterns are mostly used for visualizing hierarchical data and ``viewing relationships among disparate entities'', except for the Spiral pattern which is used to visualize ``serial periodic data'' \cite{draper2009survey}.

Concentric pattern have an ``origin at or near center of canvas'' from where rings are plotted outwards and ``each ring divided into multiple sectors''  \cite{draper2009survey}. Filelight is an example of concentric pattern and it is a ``filesystem browser based in part on the Polar TreeMap metaphor'' \cite{HowellFilelight, draper2009survey}.

Spiral pattern consists of a ``spiral-shaped glyph'' that starts from the center of the canvas \cite{draper2009survey}. Certain patterns can emerge when arranging the data according to its periodicity, as observed by Carlis and Konstan \cite{draper2009survey, carlis1998interactive}. RankSpiral is an ``interface for search engines'' developed using Spiral patterns \cite{spoerri2004rankspiral, draper2009survey}.

Euler pattern has ``multiple circles placed inside (or adjacent to) a larger circle'' often linked to represent a nested visualization of a hierarchy \cite{draper2009survey}. An example is Zoomology, which uses the ``outer ring'' as the actual root of the hierarchical structure where ``each node’s children are rendered as inner circles'' \cite{hong2003zoomology, draper2009survey}.

\paragraph{Ring}

The Ring visualization distributes the nodes ``around the circumference'' and its common use is to identify relationships between the nodes \cite{draper2009survey}. The classification divides this group into \textbf{Connected} and \textbf{Disconnected} Rings.

Connected Rings have the nodes connected by ``line segments'' and in some cases ``additional nodes''  are positioned in the ``ring’s interior'' \cite{draper2009survey}. A popular example is the Circos visualization tool for ``identification and analysis of similarities and differences arising from comparisons of genomes'' \cite{krzywinski2009circos}.

Disconnected Rings follow the same principle but the nodes have no connections between them, thus representing large datasets without the clutter that Connected Rings have when portraying the relationships between the nodes \cite{draper2009survey}. SQiRL is an example of this pattern, it is a tool that visualizes the ``opinion polls'' by breaking down the ``respondent's answers to selected questions'' placing them ``around the circumference'' \cite{draper2008votes, draper2009survey}.


In the context of health data visualization, radar visualizations have a potential to  visualize large amount of datasets due to their clarity in the data representation. However, the potential use of interactivity needs to be addressed by these visualization tools. The ability to represent relevant information should be embedded in the visualization tool leveraging from modern technologies such as Web browsers and Web services as well as with current interfaces such as touchscreens.

\begin{figure*}[tbh]
\centering
\includegraphics[scale=1]{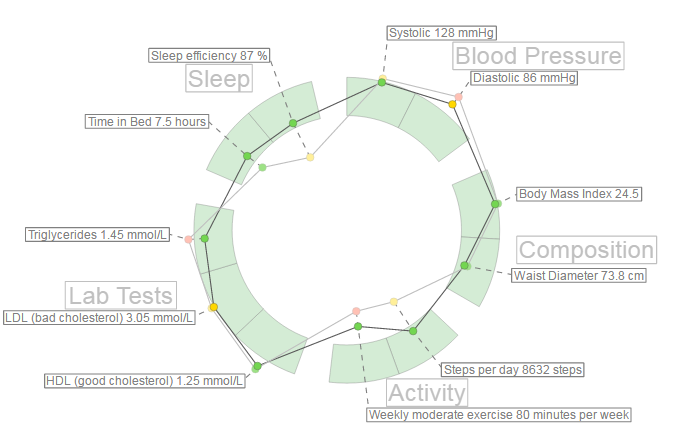}
\caption{A simple example of hFigures. The measurements are separated by groups in sectors.}
\label{figureHFiguresSimple}
\end{figure*}

\subsection{The Health Graph}

The health graph, or hGraph was developed by MITRE Corporation and released to the public in 2010 under the Apache v2.0 license. The design intention of the hGraph is to facilitate the graphical representation and understanding of health data. The data can come from a wide-range of sources such as laboratory tests, physical activity, nutrition, sleep monitors and other sources. The domain of this visualization technique includes Personal Health Informatics, EHR and Personal Health Record (PHR) visualization \cite{follett2012hgraph}.

Following the classification from Draper \textit{et al}, the hGraph could be classified as a Polar Plot and Ring, following the design patterns of a Star and a Connected Ring.

The hGraph design consists of a circular space with an area defined by to circumferences. The area represents the minimum and maximum recommended values for a given measurement. For instance, the minimum and maximum recommended fat percentage of a person in a given age. The measurements are represented as circles and their position in the circular space represents how far or close they are from the recommended values. The position is normalized according to the recommended values.

The values are distributed in a circular space. A graph is formed by joining the data points around the circular area. This polygon or graph reveals a pattern and its shape provides a quick overview of the general situation of all the values and how they deviate from the recommendations. The hGraph design highlights values outside of the recommendation by using the red color on the data points and by modifying the shape of the graph. The rationale of the hGraph is that if the same measurements are plotted in the same order for various cases, then the graph patterns can reveal similar shapes associated with certain health conditions.

\paragraph{Web-based Solution}
Web-based solutions for data visualization provide flexibility, as they can be accessed by any web browser, either from mobile devices or personal computers. The hGraph uses a web approach via HyperText Markup Language (HTML) and Scalable Vector Graphics (SVG). The programming language of the library is JavaScript and is built using the Data-Driven Documents library. Data-Driven Documents (D3.js) library provides free access to the Document Object Model (DOM), which is the substrate that enables the programmer to interface with the graphical representations in a web browser \cite{bostock2011d3}.

\paragraph{hGraph as an Insightful Visualization}
Based on the approach proposed by C. North \cite{north2006toward}, a previous study \cite{EMBCLedNiem} compared visualizations based on how well users derived meaningful insights. The study compared the hGraph visualization along with four alternatives based on the Graphical Perception Framework proposed by Cleveland McGill \cite{cleveland1984graphical}. The study compared the same data plotted with these five alternatives plus a control group which had the numerical data with no visualization. The data was comprised by a set of measurements of two modelled patients. The first patient had an elevated at risk of developing Type II Diabetes and the second one had a low risk due to a healthy lifestyle (regular exercise and a balanced diet). The evaluation followed the insight-based methodology similar to other experiments for visualizing genetic expressions \cite{saraiya2004evaluation}. The experiment determined how these visualizations can enable users to understand the overall health situation of the modelled patient with poor health, as well as the possible causes behind that patient's situation. The hGraph was found to be the most effective solutions for creating meaningful insights and to help users to better understand the data.




Figure \ref{figureHGraphZoom} is an example of the hGraph visualization. The figure was extracted as a snapshot as the library generates an SVG document structure that cannot be exported outside the browser. We address this issue in the next section of the article. The hGraph hides the measurements when the zoom level is low, meaning that the user has zoomed out. The shape is an average of the deviation of each measurement under the same category. When the user zooms in, the details are revealed and the rest of the information becomes visible, that is the numerical values and positions (with respect to the recommendation) of the measurements.


\section{Implementation} 

In this chapter we detail the implementation methodology and the key features implemented in the library. We named the library Health Figures (hFigures) because it is based on the design principles of the hGraph. hFigures makes an emphasis on multiple graphs, or figures, in order to provide a graphical representation of evolution of the data over time (multiple snapshots of the data at certain points in time).

\subsection{Methodology}

The implementation of the hFigures followed the Extreme Programming methodology \cite{roebuck2012agile, beck2000extreme}. The main key requirement was to provide a visualization which represents the changes in the overall health situation. In addition, the implementation addressed the features that the users requested in a previous study \cite{EMBCLedNiem}.

Extreme Programming focuses on releasing and reviewing functional software continuously \cite{beck2000extreme}. Often these requirements change and the programming practice is to address this changes by prioritizing them at the top of the change list.

During the implementation of the library, our research group provided the continuous review process of the software. The research group has expertise in Health Sciences, Signal Processing, User Design, Software Engineering and Machine Learning. Requirements often changed and new releases were assessed by the group. The development of the health monitoring application followed the practice of pair programming, as it is often the case in Extreme Programming \cite{roebuck2012agile}.

\subsection{Data Source}

\begin{figure}[tbh]
\centering
\includegraphics[scale=0.55]{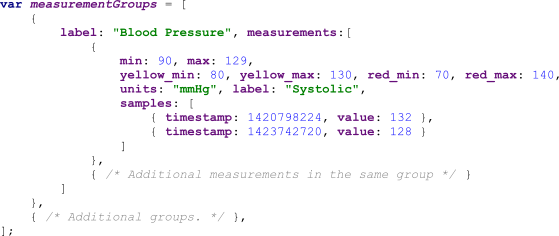}
\caption{JSON data source file. The data source file structured as a JSON file.}
\label{codeDataSource}
\end{figure}

We use a JSON (JavaScript Object Notation) format to read the data, in which the measurements are grouped according to their categories. The groups contain an array of samples, which represent the values obtained from a measurement (steps per day, cholesterol, triglycerides, blood sugar or depression level using \cite{poutanen2010validity}). The samples contain a timestamp in Unix Epoch format and the value of the measurement. The Unix Epoch format is the number of seconds since the first of January 1970, Greenwich Meridian Time (GMT). An example of the data source is in figure \ref{codeDataSource}, it shows the first measurement of the group "Blood Pressure" which in this case is comprised by Systolic and Diastolic measurements and each of them have two samples taken at two particular times, Friday 9th of January 2015 at 10:10:24 GMT (1420798224 --- seconds) and Thursday 12th of February 2015 at 12:05:20 GMT (1423742720 --- seconds).

\subsection{SVG Document Export}

The SVG document structure we designed in our implementation can be exported to a separate file outside the web browser. The short-term objective is to build a tool-kit that enables researchers to visualize their data with our implementation so they can use the generated SVG file in articles, posters, presentations or other applications. For instance, figure \ref{figureHFiguresSimple} has been exported as an SVG document from the browser into this article. SVG export is possible due to the rendering of our algorithm which does not depend on JavaScript or Cascade Style Sheet (CSS) styling properties to produce a finalized document. The library build the entire image as a stand-alone document. The hGraph library unfortunately does not produce a complete document but instead depends on CSS and JavaScript code to make the image visible.

\subsection{Constant Graph Shape}

Figure \ref{figureHGraphZoom} shows an hGraph example and figure \ref{figureHFiguresSimple} shows an hFigures example. The hGraph computes the average of the deviation of the measurements in order to show the polygon or graph, as a representation of the overall health assessment. However, some measurement might deviate towards a lower value and while others towards a higher one, thus the average position would be roughly the middle recommended area. For this reason, hFigures does not change the shape of the graph if the user zooms in or out. Showing and hiding the measurement labels is the only reaction to the zooming events from the user at the moment. This avoids clutter when the user wished to have a quick glance at the picture but keeps the graph with the same shape.

\subsection{Layout Construction}

\begin{figure}[tbh]
\centering
\includegraphics[scale=0.55]{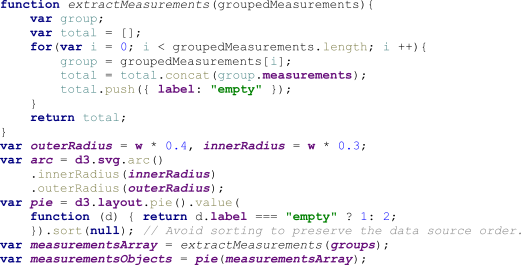}
\caption{Layout construction code. The layout for distributing the measurements is build using d3 pie layout leaving spaced between the measurement groups.}
\label{codeLayoutConstruction}
\end{figure}

The measurement groups are represented using a circular layout divided in sectors. The goal is to have a clear division between the groups as they represent the different aspects of the overall health. hFigures uses the d3 pie chart layout and modifies the data source provided to the layout. All the measurements have the same numerical value and at the end of the group, we insert an extra value in order to leave a blank space between the circular area sectors. The result is visible in the hFigures example shown in figure \ref{figureHFiguresSimple} and the code that produced this visualization is in figure \ref{codeLayoutConstruction}. The pie layout constructs the sectors of the circular area based on a data source. When we provide an array of numbers, the layout uses the numbers to calculate the proportions of the area. In order to achieve the layout construction that we have designed, the array has the same constant number multiple times, the number of measurements plus an additional number for each group. 

\subsection{Color-coded Entries}

The data source can contain additional sets of value ranges. For instance a warning range of values can let the users know when a value has reached a level that requires attention but has not yet reached a critical point. We followed the users' feedback that recommends a traffic light-based approach. The green color means that the values are within the recommended, yellow suggest a warning or follow-up action needed and the red indicated a critical threshold has been passed.

\begin{figure}[tbh]
\centering
\includegraphics[scale=0.55]{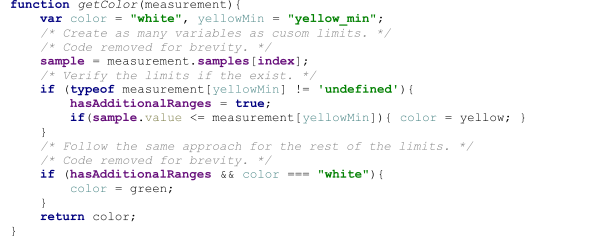}
\caption{JavaScript code to determine if additional ranges are provided. In JavaScript the data source could contain additional ranges, these are properties in an object that need to be checked beforehand and if the exists, compare the values accordingly.}
\label{codeColorCoded}
\end{figure}

In the implementation, the library verifies if these additional ranges are present in the measurement definition. In order to verify if the property of the object exists, JavaScript provides a qualifier method, \texttt{typeof}. The returned value must be compared with the keyword definition for properties that are not present in an object, the keyword \texttt{undefined} has been suggested by Mozilla Developer Network \cite{MDNJS}, a highly reputable source for Web development. The code is shown in figure \ref{codeColorCoded}.

\subsection{Multiple Graph}

The dataset is structured as a set of measurements where each has its own collection of samples. In order to compare the evolution of these measurements, the hFigures library allows the graphical representation of any number of samples. The result is a set of graphs or polygons overlapping or stacking with each other. In order to differentiate them, we use a lighter set of colors so that the users can see the difference between two points in time. As an example, figure \ref{figureHFiguresSimple} shows two different samples for each measurement. This example portrays a modeled person that has been active in a health coach program. Some measurements have improved and are closer to the recommendation. Users repeatedly expressed that it would be very helpful to visualize two or more different points in time so as to compare how the person has evolved. 

Including multiple graphs has implications in the structure and procedures of the visualization construction. For instance, we structured the SVG document such that each measurement includes one or many plotted circles that map to each sample. The measurement labels need to be positioned considering that a plotted circles can (and probably will) overlap. This is challenge that we address in the next section by finding an optimal label space distribution to avoid labels from overlapping and also to reduce the clutter in the visualization space.

\subsection{Label Space Distribution}

After the measurements are plotted, the labels are added to increase readability. The position of the label needs to be defined within a given range to avoid overlaps and clutter.

\begin{figure}[tbh]
\centering
\includegraphics[scale=0.55]{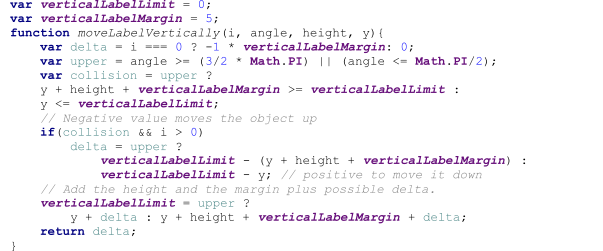}
\caption{JavaScript code for label positioning. The function in JavaScript distributed the positioning of the label to avoid overlapping and clutter.}
\label{codeVerticalLabelDistribution}
\end{figure}

\begin{figure}[tbh]
\centering
\includegraphics[scale=0.5]{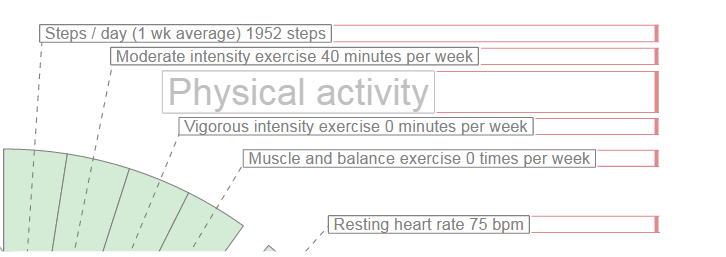}
\caption{Label positioning distribution. The labels are positioned according the their height and margin, as a result the labels do not overlap and clutter is avoided.}
\label{figureVerticalLabelDistribution}
\end{figure}

Labels need to avoid overlapping with each other and with their measurements. To solve the label overlapping problem we implemented an algorithm that starts by ordering the labels by their angular position, that is the angle at which the measurement is positioned. The next step is to calculate the height of the label and position it over the previous one in the direction that goes from the center of the visualization area upwards or downwards (depending on the angle). The idea is to begin with the center of the area, either to the left or to the right of the circles, then we work our way up or down drawing the labels into the SVG document. We add the labels as SVG elements and the use the \texttt{transform} property to position them in the corresponding place. Figure \ref{figureVerticalLabelDistribution} shows the spacing between the labels using the algorithm when drawing the labels from the center to the upper right corner. For each of the four quadrants, the library calls the method shown in figure \ref{codeVerticalLabelDistribution} which computed the position of the label as we described.

\begin{figure}[tbh]
\centering
\includegraphics[scale=0.5]{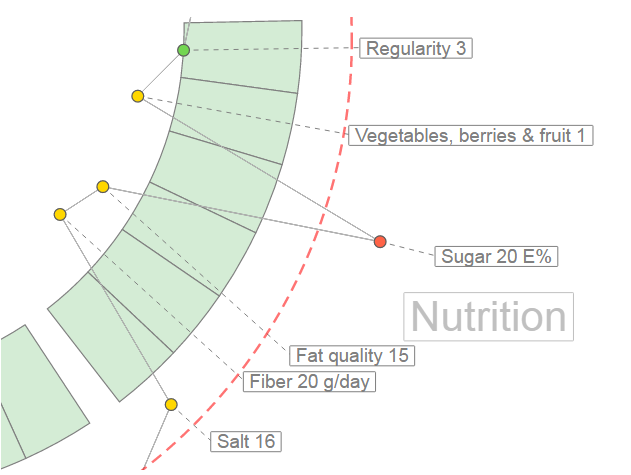}
\caption{Label positioning avoiding overlapping with measurements. Some measurements can be positioned outside of the recommended range, the labels are adjusted to avoid overlapping.}
\label{figureRadiusMeasurementDistrbution}
\end{figure}

As mentioned before, labels can also overlap with measurement circles. To avoid this problem we calculate the maximum radius from the center of the visualization area to the highest value of a measurement sample. From that starting point, we place the label in that position. In other words, for each measurement, we find the largest value of the samples. Figure \ref{figureRadiusMeasurementDistrbution} shows a sector of the hFigures where the sugar measurement label has been pushed out for a few pixels in order to avoid overlapping it with the red circle. The rest of the labels adjust to that position by leaving a user-defined margin.

\begin{equation} \label{eq:maxRadius}
	r_{label} = max(\bigcup_{i}^{n} \left \{ r_{i} \right \} ) + \textit{margin}
\end{equation}

The radius for the label is the maximum value of the samples translated as graphical coordinates plus a margin. Equation \ref{eq:maxRadius} obtains the label radius $r_{label}$ given the radii of the samples of a measurement plus the default margin $m$.





\subsection{Feature Implementation Summary}

The key improvement of hFigures is the addition of multiple graphs as a mechanism to compare the values of the health measurements over time.

The immutable shape of the graph presents the same information (values of the measurements in respect to the recommended target) regardless of zooming. This feature shows the data ``as is'' without calculating average, mean or deviation. Users stressed the importance of graphically representing the information without any calculations such as mean or accumulated values. The users expressed that showing the measurement values in hFigures helped them to derive valuable insights with just a quick glance at the data, for instance they identified measurement that fall outside of the recommended range easier and without requiring them to zoom in or out.

The hFigures library does not calculate an overall score since the users considered that this task should be the sole responsibility of a health care professional. An overall score also depends on each person under a case by case basis. For instance, the hGraph allows the user to assign weights to each measurements' value, the score is then calculated summing the value of the weights times the measurement's deviation from the recommendation. The users participating in the design process of our library expressed that an overall score would complicate the integration of the library into daily health care processes as specialists would need to review case by case to find the adequate score formula, which means specifying the weights (importance) of each measurement for a given person.

Measurements in the hFigures library are grouped in sectors which represent the category they belong to. Grouped measurements allow a clear division of categories resulting in enhanced understanding on how certain areas of wellness have changed and how, if any, they affect each other. The sectors remain visible regardless of the zoom level, users expressed that this feature provides an informative approach as the categories are always showing to which category the measurements belong to. Hiding the category labels and displaying the measurements without divisions would complicate understanding the status of health categories, such as sleep, nutrition, physical activity and others. Users expressed their confusion when they were unable to determine when a category starts or ends after zooming in and out of the hGraph.

The possibility to export the generated figure as an SVG file, allows the integration into research articles, presentations, websites, posters and other Software applications to further enhance the utility of the hFigures.

\begin{figure*}[tbh]
\centering
\includegraphics[scale=0.65]{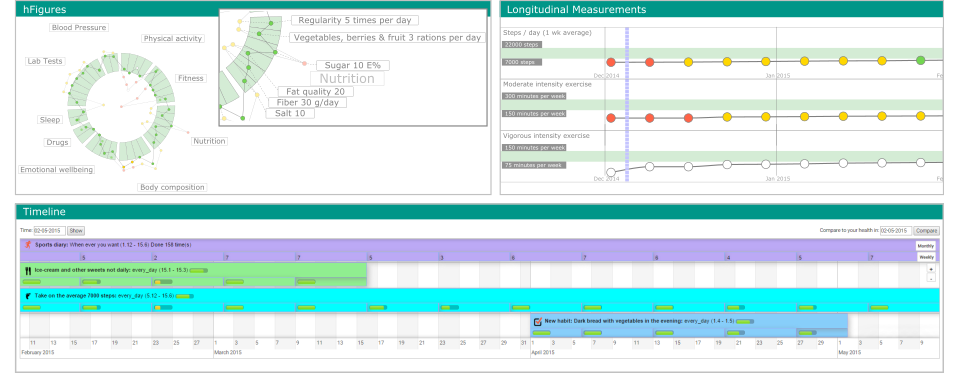}
\caption{Health Monitoring Application. The health monitoring application with the three components: hFigures, Activity Timeline and Longitudinal Measurements.}
\label{figureHealthCoachingApp}
\end{figure*}

\section{Evaluation}

The health data visualization library was placed in the context of a full application. We tested the library in a contextualized scenario where the users conducted a series of tasks and answered usability questionnaires. In this section we present the methods we used for recruiting the participants and for the usability testing of the library. We also explain the metrics measured and the rationale behind the selection of the usability testing methods.

Nielsen suggests that ```usability has multiple components and is traditionally associated with the five usability attributes, which are learnability, efficiency, memorability, errors, and satisfaction'' \cite{nielsen1994usability}. In order to assess the usability of the software solution, multiple alternatives exist in industry and research. Johnson \textit{et al}. developed a toolkit for usability testing of Electronic Health Records commissioned by the Agency for Healthcare Research and Quality of the U.S. Department of Health and Human Services \cite{johnson2011ehr}. The toolkit is built on the basis of the assessment of existing usability methods in the context of Electronic Health Records and Health Information systems. The toolkit is a detailed analysis of the usability methods, their advantages, disadvantages and appropriateness ranking.

We selected the Usability Questionnaires since it has a high appropriateness ranking \cite{johnson2011ehr}. We were able to recruit three usability experts to conduct the Heuristic Evaluation and the Cognitive Walkthrough, both are recommended techniques to complement the evaluation. We concluded the evaluation using the principles of Controlled User Testing.

\subsection{Continuous Health Monitoring Application}

In order to test the hFigures library, we designed an application for visualizing the health situation of a modeled patient and how this has changed over time within a health coaching program. The objective is to help the users in the decision making process of assessing the overall health situation and whether or not the health program has provided benefits.

The application has three components: activity timeline, the hFigures data visualization library and longitudinal measurements. Figure \ref{figureHealthCoachingApp} shows the three components. The figure is not a screenshot of the application but rather an extraction of the SVG documents embedded in the HTML file, except for the timeline.

\paragraph{Activity Timeline} This component represents the health interventions (particular actions) that the modeled patient has done during the health coaching program. During the program, several snapshots of the patient's overall health are taken and visualized using the hFigures library.

\paragraph{hFigures} The developed hFigures library is utilized to display the set of measurements taken during the health coaching program. These measurements describe an overview of the health situation of the modeled patient. Users can change the time at which the snapshot was taken to compare changes over time as a result of the health interventions.

\paragraph*{Longitudinal Measurements} The application also displays the same set of measurements using longitudinal temporal representation. We included this component to provide additional details and trends on how the measurements have changed over the coaching program.

\subsection{Heuristic Evaluation}

Heuristic Evaluation requires at least one expert in the area of human-computer interaction \cite{nielsen1994usability, johnson2011ehr}. For our evaluation we recruited three experts, and they assessed the application using Nielsen's heuristics \cite{nielsen1994usability}. The evaluation has 11 metrics evaluated using a seven point Likert scale, where the value 1 indicates ``strongly disagree'' and 7 ``strongly agree''


Heuristics are ``rules of thumb'' comprised of 10 principles meant to assist the Human-Computer Interaction specialist in the usability assessment \cite{johnson2011ehr, Nielsen10UsabilityHeuristics}. We explain the principles of the Heuristic Evaluation according Nielsen \cite{Nielsen10UsabilityHeuristics}.
\begin{enumerate}
\item \textit{Visibility of the System Status}: Refers to continuous feedback on the status of the system ``within reasonable time'' (Feedback).
\item \textit{Match between system and the real world}: The use of the language should be familiar to the user so that conversations follow a ``natural and logical order'' avoiding technical terminology unfamiliar to the intended user audience (Speak the User's Language).
\item \textit{User control and freedom} : Allow the user to recover from erroneous navigational options with ``clearly marked'' access options (Clearly Marked Exits).
\item \textit{Consistency and standards} : Follow the same language and terminology to avoid the user from guessing the meaning of ``words, situations, or actions''(Consistency).
\item \textit{Error prevention} : Avoid ``error-prone'' options in the system whenever possible and for those cases when the problematic options cannot be avoided, then present the user confirmation dialogues (Prevent Errors).
\item \textit{Recognition rather than recall} : Present visible options to the user at all times so as to avoid the effort of remembering previously stated instructions. Whenever options cannot be visible, make them ``easily retrievable whenever appropriate'' (Minimize User Memory Load).
\item \textit{Flexibility and efficiency of use} : The interface should accommodate the novice and advance user by providing ``tailored frequent actions'' (Shortcuts).
\item \textit{Aesthetic and minimalist design} : The dialogues should only contain relevant and clear information that is timely needed at that particular state of the interface (Simple and Natural Dialogue).
\item \textit{Help Users recognize, diagnose, and recover from errors} : Plain language should be used in error messages, and whenever possible they should provide helpful information so that the users can take constructive actions. (Good Error Messages)
\item \textit{Help and documentation} : Some systems require documentation and guidelines to explain briefly how to accomplish specific tasks in concrete steps.
\end{enumerate}

\subsection{Cognitive Walkthrough}

Wharton \textit{et al}. developed the Cognitive Walkthrough for usability testing \cite{wharton1994cognitive}. Johnson \textit{et al}. summarize this method as a ``usability inspection method that compares the users’ and designers’ conceptual model and can identify numerous problems within an interface'' \cite{johnson2011ehr, wharton1994cognitive}.

Cognitive Walkthrough has successfully been used to evaluate usability of Healthcare Information Systems \cite{johnson2011ehr, peute2007significance, karahoca2010information, cohen2004cognitive, khajouei2009usability} and Web Information Systems \cite{blackmon2002cognitive}.

Since Cognitive Walkthroughs ``tend to find more severe problems'' \cite{sears1997heuristic, johnson2011ehr} but ``fewer problems than a Heuristic Evaluation'' \cite{beuscart2007human, johnson2011ehr} we included both methods in our evaluation.

\subsection{Laboratory Testing}

Regarded as the ``golden standard'' for usability testing \cite{newman199810}, Laboratory Testing collects ``qualitative and quantitative'' data ``since it collects both objective data such as performance metrics (e.g., time to accomplish the task, number of key strokes, errors, and severity of errors) and subjective data such as the vocalizations of users thinking aloud as they work through representative tasks or scenarios'' \cite{johnson2011ehr}. 

Controlled user testing is comprised of ``a series of commonly used task scenarios'' where users are asked to conduct these tasks using a ``thinking aloud'' \cite{johnson2011ehr, ericsson1980verbal, nielsen1994usability}. This process requires ``users to talk aloud about what they are doing and thinking'' while they complete the tasks using the system \cite{johnson2011ehr, ericsson1980verbal, nielsen1994usability}.

As the ``golden standard'' in usability testing, this method has been widely used in evaluating Health Information Systems \cite{johnson2011ehr, borycki2009novice, currie2003clinical, hasman2006development, wu2008usability}

The data exploration tasks are designed to assist the decision making process on the health situation of the modeled patient. The usability scenario was the main goal of the intended use of the application. We explain the participants the purpose of the application, which is to facilitate the decision making process weather or not the overall health situation of the modeled patient is favourable and weather or not the health coaching program was beneficial for the patient. The tasks are designed to represent the common usage of the application, namely to find the measurements inside and outside of the recommendation and to identify the areas that improved and need even further improvement. The tasks given to the participants are shown in the following list.
\begin{enumerate}
\item How many areas of health are displayed in the hFigures?
\item Choose one of these areas and point to its measurements.
\item Identify one measurement inside the recommended values and another one outside.
\item Identify the measurement that is the furthest from the recommended values.
\item What does the green, yellow and red circles mean?
\item Has the overall health improved after coaching?
\item Which area of health has improved the most after health coaching?
\item Which measurements show the biggest improvement?
\item Understand the difference between the points inside and outside the recommended area.
\end{enumerate}

\subsection{Usability Questionnaires}

We followed the recommendations from Johnson \textit{et al}. and used this method in our evaluation. Usability Questionnaires are ``the most common'' method to ``collect self-reported data'' from the ``users’ experience and perceptions after using the system in question'' \cite{johnson2011ehr}. Although the data collected is self-reported, some questionnaires have reliability in measuring several usability metrics such as ``satisfaction, efficiency, effectiveness, learnability, perceived usefulness, ease of use, information quality, and interface quality'' \cite{johnson2011ehr}.

We used two Usability Questionnaires to evaluate the usability of our application, Computer System Usability Questionnaire (CSUQ) and After Scenario Questionnaire (ASQ) \cite{lewis1995ibm}. Table \ref{tbl:questionnaires} shows the length, reliability and the metrics of the questionnaires. These questionnaires use a seven-point Likert scale from ``strongly disagree'' up to ``strongly agree''.

\begin{table}[h!]
\caption{Standard Questionnaires Table. The table lists the standard questionnaires we used for the user evaluation of the system with their length, reliability and metrics.}
      \begin{tabular}{lccL{3.5cm}}
        \hline
         & Items & Reliability & Metrics \\ \hline
        \multirow{4}{*}{CSUQ} & \multirow{4}{*}{19} & 0.93 & Usefulness \\
         & & 0.91 & Information Quality \\
   		 & & 0.89 & Interface Quality \\
   		 & & 0.95 & Overall Usability \\ \hline
        \multirow{3}{*}{ASQ} & \multirow{3}{*}{3} & \multirow{3}{*}{0.93} & Ease of Task Completion \\
         & & & Time Required to Complete the Task \\
         & & & Satisfaction \\ \hline
		
      \end{tabular}
\label{tbl:questionnaires}
\end{table}

\paragraph{Computer System Usability Questionnaire (CSUQ)} The questionnaire was developed by IBM and it is a ``slight'' modification of the Post-Study System Usability Questionnaire (PSSUQ) \cite{lewis1992psychometric}. Table \ref{tbl:questionnaires} shows the reliability of this questionnaire. The questionnaire has high `` coefficient alpha'' with a reliability 0.95 in total and ``0.93 for system usefulness, 0.91 for informational quality, and 0.89 for interface quality''  \cite{johnson2011ehr, lewis1995ibm, lewis1992psychometric}. We selected this questionnaire since it has been successfully used in the Healthcare domain \cite{johnson2011ehr, jaspers2008pre} and in the evaluation of ``of a guideline-based decision support system'' \cite{johnson2011ehr, goud2008subjective}.

\paragraph{After Scenario Questionnaire (ASQ)} An additional questionnaire developed by IBM \cite{johnson2011ehr, lewis1995ibm, lewis1991psychometric} and designed to measure the user satisfaction after scenario usability studies have been completed \cite{johnson2011ehr, lewis1992psychometric, bangor2008empirical}. This questionnaire measures the ``ease of task completion, time required to complete the tasks, and satisfaction with support information'' \cite{johnson2011ehr}. Since we already designed the scenario for the evaluation of the system, we included this questionnaire in our study.

\subsection{Data Model}

Similar to the study we conducted in the insight-based methodology \cite{EMBCLedNiem}, we modeled a patient using clinical expertise of a physician along with the most common symptoms for developing Type II Diabetes. The modeled patient consisted of a set of measurements over time comprised of the following parameters:

\begin{itemize}
	\item Blood pressure: systolic and diastolic blood pressure 
    \item Physical activity: weekly active days\cite{us2008physical, physical2008physical}, steps per day \cite{tudor2004many} 
    \item Body composition: Body Mass Index (BMI), waist diameter and fat percentage
    \item Sleep: time in bed, time asleep  
    \item Fitness: resting heart rate, fitness index \cite{oja2013tester,laukkanen1992validity}, muscular force, muscular endurance and balance \cite{suni2009fitness}
    \item Lab Tests: hemoglobin, fB-Gluc, cholesterol, HDL, LDL, triglycerides 
    \item Nutrition: meal regularity, type of meals (vegetables and fruits), sugar intake, fat quality, fiber and salt intake.
    \item Drugs: tobacco (cigarettes per day), alcohol abuse, drug abuse (narcotics), medication abuse
    \item Emotional wellbeing: depression level \cite{poutanen2010validity}, stress level and stress recovery \cite{firstbeat2014, teisala2014associations} and optimism \cite{scheier1994distinguishing}.
\end{itemize}	

\subsection{Recruitment}

We recruited a total of 14 participants following similar usability studies and Faulkner's \cite{faulkner2003beyond} recommendation of conducting usability tests with 10 to 20 users ``in order to find 90 to 95\% of usability problems'' \cite{berry2015usability}. Among the 14 participants we were able to recruit 3 usability experts, following the recommendations from Nielsen and a number of previous studies stating that 3 to 5 experts are needed to conduct the Heuristics Evaluation \cite{johnson2011ehr, zhang2003using, nielsen1994usability, shneiderman1992designing, molich1990improving, Tognazzini2014principles}.

Participants were recruited through the university's student email lists, self-study groups, lectures and workshops. After completing the usability tests, the participants received a movie ticket.

\subsection{Ethics}

The study we conducted was a usability evaluation using simulated data not belonging to a real person. The results of the usability tests were kept anonymous and the collected data does not include sensitive information from the participants. According to the ethical principles applied by the Finnish Advisory Board on Research Integrity, our study did not need ethics approval \cite{tenk2009}.

The experimental procedures described in this paper complied with the principles of Helsinki Declaration of 1975, as revised in 2000. All subjects gave informed consent to participate and they had a right to withdraw from the study at any time. The informed consent also explained that their names and identities will be kept confidential, that the results will not be linked to their identities, the sessions will be recorded using a Web camera and microphone for further study and that the clinical data visualized did not belong to a real person.

\subsection{Experiment Protocol}


The testing process started with the signature of an informed consent where we explained the participants the purpose of the test. Afterwards we proceed to explain a usability scenario and the tasks that the participants were asked to complete. The participants were allowed to ask questions at any time. After performing the tasks we asked the participants to fill in the Usability Questionnaires. We close the session with a briefing interview where we asked the users what they liked and disliked about the application as well as what were their recommendations for further improvements. The sessions were recorded for further study and to find the correct timing of the task completion.

\subsection{Materials and Tools} 

We conducted the usability tests in our laboratory. We used a computer with a local HTTP server running our server application and Google Chrome as the browser running our front-end application. The computer was a laptop with a camera and microphone which were used to record the session for later study. The computer was connected to a 23 inch display and a separate keyboard and mouse. The usability questionnaires were filled out using the Web portal developed by Perlman and available at the following address \url{http://garyperlman.com/quest/} .

\section{Results}

\subsection{Heuristic Evaluation}

The three expert users answered the Heuristic Questionnaire in order to identify problems with the user interface of the health monitoring application. The three experts agreed and in some cases strongly agreed with most of the indicators. One expert found the instructions for adjusting the time of the visualization tool to be demanding. The expert addressed this comment to the integration interface that allows the time to be adjusted and thus visualized. The remark was not addressed to the graphical representation of the data using hFigures. The results of the evaluation are summarized in table \ref{tbl:heuristicsResults}. The average response was 6.3 out of 7 points.

\begin{table}[h!]
\caption{Heuristic Evaluation Results. The table summarizes the results of the Heuristic Evaluation conducted by three usability experts.}
      \begin{tabular}{p{3.85cm}R{1.2cm}R{1.2cm}}
      \hline
      Heuristic & Average Response & Standard Deviation \\ \hline 
      Visibility of system status & 6.00 & 1.00 \\
      Match between system and the real world & 6.33 & 0.57 \\
      User control and freedom & 6.33 & 0.57 \\
      Consistency and standards & 6.67 & 0.57 \\
      Error prevention & 6.33 & 0.57 \\
      Recognition rather than recall & 4.67 & 0.57 \\
      Flexibility and efficiency of use & 6.67 & 0.57 \\
      Aesthetic and minimalist design & 7.00 & 0.00 \\
      Help users recognize, diagnose, and recover from errors & 6.33 & 0.57 \\
      Help and documentation & 6.67 & 0.57 \\ \hline
      Nielsen heuristic evaluation & 6.30 & 0.56 \\ \hline
      \end{tabular}
\label{tbl:heuristicsResults}
\end{table}

\subsection{Cognitive Walkthrough}


During the Cognitive Walkthrough, the concept of the health monitoring application was explained to the usability experts. The purpose of the application was explained in the context of the health situation of the modeled patient and how the application visualizes the changes in the health situation over time. We used the usability scenario and tasks to confirm that the interface supports the intended use of the application. The questions comprising the walkthrough, as described by Wharton \textit{et al} \cite{wharton1994cognitive}., were correctly answered by the expert users thus no design or mismatch errors were found.

\subsection{Controlled User Testing}

\begin{table}[h]
\caption{Controlled User Testing Results. The table summarizes the results of the 14 users performing the 9 tasks.}
      \begin{tabular}{lC{1.5cm}C{0.75cm}R{1.2cm}R{1.2cm}}
      \hline
      Task & Successfully Completed & Errors & Average Time (seconds) & Standard Deviation (seconds)\\ \hline
      Task 1 & 14 & 4 & 12.21	&	12.60 \\
      Task 2 & 11 & 0 & 10.00	&	12.38 \\
      Task 3 & 14 & 1 & 10.78 &  5.38 \\
      Task 4 & 14 & 2 &  6.78	&  4.98 \\
      Task 5 & 14 & 0 & 17.50	&  6.60 \\
      Task 6 & 14 & 2 & 16.07	& 16.52 \\
      Task 7 & 14 & 0 &  6.21	& 5.591 \\
      Task 8 & 14 & 1 &  7.85	& 5.882 \\
      Task 9 & 13 & 1 &  9.23	& 4.729 \\ \hline
      \end{tabular}
\label{tbl:tasksResults}
\end{table}

Table \ref{tbl:tasksResults} summarizes the results of the completed, number of errors, average time to complete the task and the standard deviation. All participants completed 7 of the 9 tasks. Task 2 was the most problematic, we asked users to ``choose one of these areas and point to its measurements'', we found that 3 participants were not able to understand the task thus unable to complete it. The second most problematic task was number 9, ``understand the difference between the points inside and outside the recommended area'', where one participant was unable to complete it successfully incurring in one non-crucial mistake (an error that prevent the task to be completed). 

Additional non-crucial errors occurred in tasks 1, 3, 4, 6 and 8. The large number occurred in the first task due to the initial values set in the default zoom level of the hFigures component. After the usability testing, we corrected this problem by adjusting the initial zoom level to include the whole figures inside the container.

\subsection{Usability Questionnaires}

\paragraph{Computer System Usability Questionnaire (CSUQ)}
We computed the results according to Lewis, obtaining the average of ``items 1 through 19'' to determine the overall usability rating of the system. System usefulness is the average of items 1 to 8, information quality 9 through 15 and interface quality 16 through 18 \cite{lewis2002psychometric}.

\begin{table}[h!]
\caption{Computer System Usability Questionnaire Results for the System Usefulness assessment. The table shows the results of the questions corresponding to the System Usefulness with its average and standard deviation.}
      \begin{tabular}{L{3.85cm}R{1.2cm}R{1.2cm}}
      \hline
      Question & Average Response & Standard Deviation \\ \hline
      Overall, I am satisfied with how easy it is to use this system & 6.29 & 0.99 \\
      It was simple to use this system & 6.07 & 1.20 \\
      I can effectively complete my work using this system & 6.07 & 1.07 \\
      I am able to complete my work quickly using this system & 5.86 & 1.40 \\
      I am able to efficiently complete my work using this system & 6.21 & 0.89 \\
      I feel comfortable using this system & 6.21 & 0.97 \\
      It was easy to learn to use this system & 6.43 & 0.85 \\
      I believe I became productive quickly using this system & 5.93 & 1.26 \\ \hline
      System Usefulness & 6.13 & 0.93 \\ \hline
      \end{tabular}
\label{tbl:CSUQSystemUsefulnessResults}
\end{table}

\begin{table}[h!]
\caption{Computer System Usability Questionnaire Results for Information Quality.}
      \begin{tabular}{L{3.85cm}R{1.2cm}R{1.2cm}}
      \hline
      Question & Average Response & Standard Deviation \\ \hline
      The system gives error messages that clearly tell me how to fix problems & 4.50 & 2.44 \\
      Whenever I make a mistake using the system, I recover easily and quickly & 5.43 & 1.95 \\
      The information (such as online help, on-screen mes-sages, and other documentation) provided     with this system is clear & 5.29 & 1.90 \\
      It is easy to find the information I needed & 6.07 & 1.27 \\
      The information provided for the system is easy to un-derstand &  5.93 & 1.39 \\
      The information is effective in helping me complete the tasks and scenarios & 6.14 & 1.17 \\
      The organization of information on the system screens is clear & 6.29 & 1.14 \\ \hline
      Information Quality & 5.66 & 1.20 \\ \hline
      \end{tabular}
\label{tbl:CSUQInformationQualityResults}
\end{table}

\begin{table}[h!]
\caption{Computer System Usability Questionnaire Results for Interface Quality.}
      \begin{tabular}{L{3.85cm}R{1.2cm}R{1.2cm}}
      \hline
      Question & Average Response & Standard Deviation \\ \hline
      The interface of this system is pleasant & 6.36 & 1.00 \\
      I like using the interface of this system & 6.36 & 0.92 \\
      This system has all the functions and capabilities I expect it to have & 6.00 & 1.18 \\ \hline
      Interface Quality & 6.24 & 0.99 \\ \hline
      \end{tabular}
\label{tbl:CSUQInterfaceQualityResults}
\end{table}

\begin{table}[h!]
\caption{Computer System Usability Questionnaire Results for Overall Usability, System Usefulness, Information and Interface Quality.}
      \begin{tabular}{L{2.5cm}lR{1.2cm}R{1.2cm}}
      \hline
      Metric & Questions & Average Response & Standard Deviation \\ \hline
      Overall Usability   &  1-19 & 6.02 & 1.04 \\ 
      System Usefulness   &  1-8  & 6.13 & 0.93 \\ 
      Information Quality &  9-15 & 5.66 & 1.20 \\ 
      Interface Quality   & 16-18 & 6.24 & 0.99 \\ \hline
      \end{tabular}
\label{tbl:CSUQOverallUsabilityResults}
\end{table}

Table \ref{tbl:CSUQSystemUsefulnessResults} shows the results of the system usefulness. The system obtained an average of 6.13 out of 7. Table \ref{tbl:CSUQInformationQualityResults} shows the results of the information quality metric where the application scored a total average of 5.66. The average value is still within the ``agree'' response of the participants, however the notable low value compare to the other metrics might be due to the amount of information presented in textual format in the application. The information was encoded using graphical representations and even though a help document was included in the system, the text was not likely to fulfil the users' expectations. Table \ref{tbl:CSUQInterfaceQualityResults} shows the score for the interface quality where the application obtained an average of 6.24 out of 7. The combined results are shown in table \ref{tbl:CSUQOverallUsabilityResults}. The score of the overall usability is 6.02 with a standard deviation of 1.04. We can determine that all the participants at least ``agreed'' in the Likert scale that the application was useful for the decision making process of assessing the health situation and evolution of the modeled patient.

\paragraph{After Scenario Questionnaire (ASQ)} The average response for the ease of task completion was 6.64 with a standard deviation of 0.842 and for the time required to complete the task 6.64 and a standard deviation of 0.497. The overall satisfaction was 6.46 and a standard deviation of 0.53. The usability of the system had a high score in the ASQ results meaning that the system was suitable for the scenario in the context of the health data visualization of the modeled patient and its evolution over time.

\begin{table}[h!]
\caption{After Scenario Questionnaire Results.}
      \begin{tabular}{L{3.85cm}R{1.2cm}R{1.2cm}}
      \hline
      Question & Average Time (seconds) & Standard Deviation (seconds)\\ \hline
      Overall, I am satisfied with the ease of completing the tasks in this scenario & 6.64 & 0.84 \\
      Overall, I am satisfied with the amount of time it took to complete the tasks in this scenario & 6.64 & 0.49 \\
      Overall Satisfaction of the system & 6.46 & 0.53 \\ \hline
      \end{tabular}
\label{tbl:ASQResults}
\end{table}

\subsection{Identified Issues and Suggested Improvements}

The feedback shows that the main problem was the incomplete visibility of the hFigures in the application component window. Users also requested to show the detailed information as a hovering pop up window in the second figure (measurements before the coaching program). Currently only the latest measurements have the hovering window however users requested that both measurements (the before and after) should contain the same functionality. Additional information was needed in the measurements that contained numerical scales, such as the depression index. A more contextualized approach explaining the meaning of the values can help the user understand the measurements and thus the overall health situation of the patient better.

\begin{figure*}[tbh]
\centering
\includegraphics[scale=0.7]{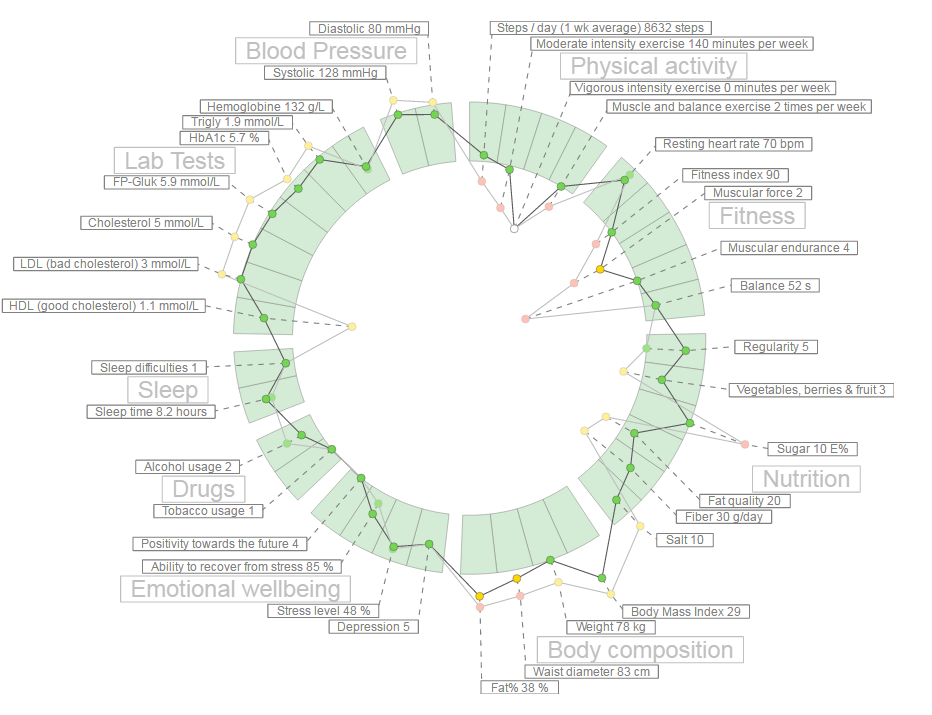}
\caption{A heterogeneous hFigures example. An overview of a modeled person comprised of several measurements with two time snapshots showing its evolution over time.}
\label{figureHFiguresComplex}
\end{figure*}

\section{Discussion}

The value of a data visualization depends on the knowledge that it can convey to the public. In this section, we claim that hFigures has the potential to be used both in the clinical and personal wellness applications. Large amounts of measurements do not clutter the visualization area as a result of our implementation, figure \ref{figureHFiguresComplex} shows an example of a large visualization of a modeled patient. The comparison of multiple graphs can provide a meaningful visualization to individuals and clinicians alike. The implementation of hFigures follows an extensible approach and even though it was designed to be used for health data visualization, any dataset that has a target range of values as a reference can be visualized.

\subsection{Translation to Health Care}

The wide variety of EHR formats and data sources from self monitoring applications comprise a challenge in unifying the data in order to provide an overview of a patient. Currently, most of the data sources contain the date when the sample was extracted, whether it is a tracking device or a blood test. This sample date already provides the timestamp required by the hFigures data source file. The values of the samples are the main object of study in a measurement, for instance the levels of cholesterol or sugar at a given time, the number of steps per day, the percentage of body fat and several others. This information can be transformed in a simple process to build the hFigures data source following the JSON structure.

hFigures is a visualization library based on Web technologies, it uses a Web browser and the rendering of the SVG is compatible with current HTML standards. Thus providing interoperability across multiple devices including tablets, smartphones, workstations or laptops is indeed feasible. The D3.js library that hFigures is built on, enables compatibility with Internet Explorer versions 8 and higher using a component named \textit{Aight} \cite{aight}. Internet Explorer 8 is prevalent in hospitals and clinics due to the restrictions in installing custom software.

\subsubsection{Patient Evolution}

Multiple graphs plotted on the layout of recommended values shows the change in the data over time. This could allow clinicians to understand the evolution of certain aspects in the health of the patient. For instance, health professionals would be able to look at the effect of a trail drug treatment over time. Possibly, the collected samples of a patient before and after starting the trial would be plotted as the multiple graphs portraying the evolution of the patient. Figure \ref{figureHFiguresComplex} shows an example of a more heterogeneous dataset. The labels for each individual measurement are usually hidden when a full zoom out is performed by the user. For the purpose of demonstrating the visualization library we have made all the labels visible.

\subsubsection{Personal Health Monitoring}

\begin{figure}[tbh]
\centering
\includegraphics[scale=0.55]{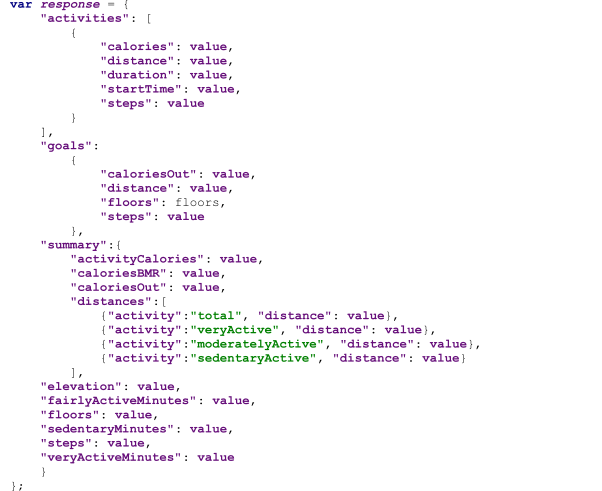}
\caption{Fitbit API example response. An example JSON response from a Fitbit activity sensor that can be transformed to a data source to be visualized by hFigures.}
\label{codeFitbitApi}
\end{figure}

As an example, Fitbit provides activity trackers and a wide-range of devices. The data collected can be obtained through their API. In most cases, providers such as Fitbit follow an HTTP REST interface. Figure \ref{codeFitbitApi} shows an example of the data Fitbit provides through its interface. The data has \texttt{"activities"} as properties of a JavaScript object. These entries have in turn a property \texttt{"startTime"} which provides the timestamp required for the hFigures data source file.

In this case, a step counter contains the property \texttt{"steps"} with the number of steps registered by the device. This and other measurement can be plotted in the hFigures following the data source file structure.

We can also use multiple graphs to show the user-defined goals as a reference in addition to the actual values of the measurement (steps per day). Figure \ref{codeFitbitApi} only show steps per day, however the documentation of the API specifies that additional data is available. This data includes distance travelled, sedentary activity, floors climbed, calories burnt and more.

Other device manufacturers such as Withings or Jawbone provide their users the possibility to use their APIs to extract collected data in a similar way. As in the case of Fitbit, all data has at least a timestamp (date, time or both) and a set of values. Recommended values for the health measurements can be obtained from healthcare professionals and public health information sources.


\subsection{Limitations}

The library reads the data in JSON format and does not support XML, which is still used by some Health Information Systems. Data files need to be included in the same HTML file and used to create an instance of the hFigures class. The library does not retrieve the data remotely by itself so the data provision is the responsibility of the developer. Other libraries provide an AJAX interface through HTTP(S) communication to an endpoint in order to retrieve the data. The information in the nodes comes entirely from data source file, which means that additional information about the measurements cannot be supplied otherwise. The extraction of the SVG file requires to export the code embedded in the HTML file. Currently no automatic export functionality is implemented.

\subsection{Further Development}

We plan to develop the library further to address the suggestions obtained from the participants of the usability testing. The next release of the library already included the fixes for the automatic adjustment of the initial zoom level to show the complete figures within the given container, usually a \texttt{<div>} element in the HTML document. The next item to address is the inclusion of additional information explaining the measurements in the hover pop up window. Additionally we need to develop an algorithm to display the values when hovering on the two figures so that the pop up windows do not overlap.

Further development contemplates a Web Service which consumes a JSON data source file and produces an SVG or a HTML document with the interaction features as a JavaScript file attached. Such service has been already requested in other projects for research purposes in order to provide a better software tool for medical decision making processes.


\section{Conclusion}




Complete and accurate visualizations of health data have been thought to empower individuals, citizens and health professionals alike, to better understand situations and take better informed decisions \cite{lesselroth2011data, shneiderman2013improving, rind2011interactive}. These decisions can be medical treatment, behaviour change practices, wellness development, health coaching program and more. In this article we detailed the underlying motivation to develop a visualization library inspired by the hGraph. 


We tested the visualization library in the context of an application by conducting usability tests comprised of Heuristics Evaluation, Cognitive Walkthrough and Usability Questionnaires. In the Heuristics Evaluation the average response was 6.3 out of 7 points and the Cognitive Walkthrough done by usability experts indicated no design or mismatch errors. In the CSUQ usability test the system obtained an average score of 6.13 out of 7, and in the ASQ test the overall satisfaction score was 6.64 out of 7. The results indicate that the library was helpful in assisting users in understanding health data and its evolution over time. 

The library is an open source tool inspired by the hGraph but with additional key improvements. However, additional improvements and fixes are needed to further develop this tool. In this article, we also discussed how this library can be used in wellness and health processes to understand the evolution of a patient's health and wellness.

Open challenges remain in studying alternative features that can help users identify relationships between measurements, visualize patterns and enable deeper exploration of the data with a higher degree of interactivity.

\section{Availability and Requirements}

\begin{itemize}
\item \textbf{Project name:} hFigures
\item \textbf{Project home page and source code repository:} \url{https://github.com/ledancs/hFigures}
\item \textbf{SciCrunch Resource ID} SCR\_014201
\item \textbf{Operating System:} Platform independent.
\item \textbf{Programming language:} JavaScript.
\item \textbf{Other requirements:} Developers willing to deploy the application need to serve the files via a Web server. Users require a Web browser to visualize the application.
\item \textbf{License:} MIT License.
\item \textbf{Any restrictions to use by non-academics:} No.
\end{itemize}

\section{Availability of Data}
The dataset supporting the conclusions of this article is available in the BioSharing repository with the identifier biodbcore-000734 at the following url: \url{https://biosharing.org/biodbcore-000734}.

The dataset is also available at Tampere University of Technology Personal Health Informatics website in the following url: \url{http://www.tut.fi/phi/?p=319}.

\section{Additional Files}

\begin{itemize}
\item \textbf{File name:} laboratory.csv.
\item \textbf{Title of data:} Results from laboratory testing.
\item \textbf{Description of data:} The data contains the task identifier, the average time to completion, number of times the task was successfully completed and the total number of errors.\newline
\end{itemize}

\begin{itemize}
\item \textbf{File name:} tasks.csv.
\item \textbf{Title of data:} Laboratory testing tasks.
\item \textbf{Description of data:} The data contains the task identifier and the instructions given to the participants to complete the task.\newline
\end{itemize}

\begin{itemize}
\item \textbf{File name:} heuristic.csv.
\item \textbf{Title of data:} Nielsen's Heuristic Evaluation.
\item \textbf{Description of data:} The data contains the results form Nielsen's Heuristic Evaluation conducted by three usability experts.\newline
\end{itemize}

\begin{itemize}
\item \textbf{File name:} csuq.csv.
\item \textbf{Title of data:} Computer System Usability Questionnaire Results.
\item \textbf{Description of data:} The data contains the results of the Computer System Usability Questionnaire answered by 14 participants. \newline
\end{itemize}

\begin{itemize}
\item \textbf{File name:} asq.csv.
\item \textbf{Title of data:} After Scenario Questionnaire Results.
\item \textbf{Description of data:} The data contains the results of the After Scenario Questionnaire answered by 14 participants.\newline
\end{itemize}



\section{Abbreviations}
    hGraph: Health graph; hFigures: Health figures; JSON: JavaScript object notation; EHR: Electronic health record; PHR: Personal health record; HTML: Hypertext markup language; SVG: Scalable vector graphics; D3: Data-driven documents; API: Application programming interface.

\section{Competing Interests}
    The authors declare that they have no competing interests.

\section{Author's Contributions}
AL developed the hFigures library. AL and HN wrote jointly this article. HN suggested several use cases that helped shaped the design process. HN contributed in the design and conception of the library. AL, HN and MAM designed the application for health coaching. AL and MAM implemented the application. AL and MAM integrated the application into a Web service. MAM conducted the user testing and analysed the results. HN and AL wrote jointly the introduction and background. AL wrote the implementation and discussion. MAM and AL wrote jointly the evaluation and results section. HN and AL wrote jointly the conclusion.
    
\section{Acknowledgements}
This research was supported jointly by TEKES (the Finnish Funding Agency for Technology and Innovation) as part of the Digital Health Revolution project, as well as the European Commission and TEKES under the ARTEMIS-JU WithMe project.
  
\section{Author's Information}
AL and HN are part of the Personal Health Informatics research group from the Department of Signal Processing of Tampere University of Technology. AL is a Software Engineer pursuing PhD studies in Health Data Visualization with a strong background in Web Engineering and Health Information Systems. HN is a PhD senior researcher with background in biomedical engineering, signal processing, electrical engineering, user interface design and service design.






\bibliographystyle{IEEEtran}
\bibliography{IEEEabrv,mybibfile}

\begin{thebibliography}{10}
\providecommand{\url}[1]{#1}
\csname url@rmstyle\endcsname
\providecommand{\newblock}{\relax}
\providecommand{\bibinfo}[2]{#2}
\providecommand\BIBentrySTDinterwordspacing{\spaceskip=0pt\relax}
\providecommand\BIBentryALTinterwordstretchfactor{4}
\providecommand\BIBentryALTinterwordspacing{\spaceskip=\fontdimen2\font plus
\BIBentryALTinterwordstretchfactor\fontdimen3\font minus
  \fontdimen4\font\relax}
\providecommand\BIBforeignlanguage[2]{{%
\expandafter\ifx\csname l@#1\endcsname\relax
\typeout{** WARNING: IEEEtran.bst: No hyphenation pattern has been}%
\typeout{** loaded for the language `#1'. Using the pattern for}%
\typeout{** the default language instead.}%
\else
\language=\csname l@#1\endcsname
\fi
#2}}

\bibitem{Hood2012613}
\BIBentryALTinterwordspacing
L.~Hood and M.~Flores, ``A personal view on systems medicine and the emergence
  of proactive \{P4\} medicine: predictive, preventive, personalized and
  participatory,'' \emph{New Biotechnology}, vol.~29, no.~6, pp. 613 -- 624,
  2012, molecular Diagnostics and Personalised Medicine. [Online]. Available:
  \url{http://www.sciencedirect.com/science/article/pii/S1871678412000477}
\BIBentrySTDinterwordspacing

\bibitem{lesselroth2011data}
B.~J. Lesselroth and D.~S. Pieczkiewicz, ``Data visualization strategies for
  the electronic health record,'' in \emph{Advances in Medicine and Biology},
  L.~V. Berhardt, Ed.\hskip 1em plus 0.5em minus 0.4em\relax New York: Nova
  Science Publishers, Inc, 2012, vol.~16, pp. 107--140.

\bibitem{follett2012hgraph}
J.~Follett and J.~Sonin, ``{hGraph: An Open System for Visualizing Personal
  Health Metrics},'' \url{http://hgraph.org/docs/hGraph_Whitepaper.pdf},
  Involution Studios, Arlington, Massachusetts, Tech. Rep., April 2012,
  accessed 25 Mar 2015.

\bibitem{EMBCLedNiem}
A.~Ledesma, H.~Nieminen, P.~Valve, M.~Ermes, H.~Jimison, and M.~Pavel, ``The
  shape of health: A comparison of five alternative ways of visualizing
  personal health and wellbeing,'' in \emph{Engineering in Medicine and Biology
  Society (EMBC), 2015 37th Annual International Conference of the IEEE}, Aug
  2015.

\bibitem{bui2007information}
A.~Bui, D.~Aberle, and H.~Kangarloo, ``Timeline: Visualizing integrated patient
  records,'' \emph{Information Technology in Biomedicine, IEEE Transactions
  on}, vol.~11, no.~4, pp. 462--473, July 2007.

\bibitem{mHealth2014}
\emph{{mHealth App Developer Economics 2014}}, research2guidance, Berlin, 2014.

\bibitem{goetz2015}
\BIBentryALTinterwordspacing
T.~Goetz. It's time to redesign medical data. [Online]. Available:
  \url{http://www.ted.com/talks/thomas\_goetz\_it\_s\_time\_to\_redesign\_medical\_data?language=en}
\BIBentrySTDinterwordspacing

\bibitem{7072838}
M.~Idris, S.~Hussain, M.~Ahmad, and S.~Lee, ``Big data service engine (bise):
  Integration of big data technologies for human centric wellness data,'' in
  \emph{Big Data and Smart Computing (BigComp), 2015 International Conference
  on}, Feb 2015, pp. 244--248.

\bibitem{few2006information}
S.~Few, \emph{Information dashboard design}.\hskip 1em plus 0.5em minus
  0.4em\relax North Sebastopol, California: O'Reilly, 2006.

\bibitem{Soomlek:2013:AWI:2532685.2532686}
\BIBentryALTinterwordspacing
C.~Soomlek and L.~Benedicenti, ``An agent-based wellness indicator:
  Experimental results and future directions,'' \emph{J. Inf. Technol. Res.},
  vol.~6, no.~2, pp. 1--23, Apr. 2013. [Online]. Available:
  \url{http://dx.doi.org/10.4018/jitr.2013040101}
\BIBentrySTDinterwordspacing

\bibitem{de2013web}
M.~de~Ridder, L.~Constantinescu, L.~Bi, Y.~H. Jung, A.~Kumar, J.~Kim, D.~D.
  Feng, and M.~Fulham, ``A web-based medical multimedia visualisation interface
  for personal health records,'' in \emph{Computer-Based Medical Systems
  (CBMS), 2013 IEEE 26th International Symposium on}.\hskip 1em plus 0.5em
  minus 0.4em\relax IEEE, 2013, pp. 191--196.

\bibitem{shneiderman2013improving}
B.~Shneiderman, C.~Plaisant, and B.~W. Hesse, ``Improving health and healthcare
  with interactive visualization methods,'' Citeseer, Tech. Rep., 2013.

\bibitem{baird1978fundamentals}
J.~C. Baird and E.~J. Noma, \emph{Fundamentals of scaling and
  psychophysics}.\hskip 1em plus 0.5em minus 0.4em\relax New Jersey: John Wiley
  \& Sons Canada, Limited, 1978.

\bibitem{cleveland1984graphical}
W.~Cleveland and R.~McGill, ``Graphical perception: Theory, experimentation,
  and application to the development of graphical methods,'' \emph{Journal of
  the American statistical association}, vol.~79, no. 387, pp. 531--554, 1984.

\bibitem{hoffman1997dna}
P.~Hoffman, G.~Grinstein, K.~Marx, I.~Grosse, and E.~Stanley, ``Dna visual and
  analytic data mining,'' in \emph{Visualization'97., Proceedings}.\hskip 1em
  plus 0.5em minus 0.4em\relax IEEE, 1997, pp. 437--441.

\bibitem{draper2009survey}
G.~M. Draper, Y.~Livnat, and R.~F. Riesenfeld, ``A survey of radial methods for
  information visualization,'' \emph{Visualization and Computer Graphics, IEEE
  Transactions on}, vol.~15, no.~5, pp. 759--776, 2009.

\bibitem{jankun2003moiregraphs}
T.~Jankun-Kelly and K.-L. Ma, ``Moiregraphs: Radial focus+ context
  visualization and interaction for graphs with visual nodes,'' in
  \emph{Information Visualization, 2003. INFOVIS 2003. IEEE Symposium
  on}.\hskip 1em plus 0.5em minus 0.4em\relax IEEE, 2003, pp. 59--66.

\bibitem{lamping1996hyperbolic}
J.~Lamping and R.~Rao, ``The hyperbolic browser: A focus+ context technique for
  visualizing large hierarchies,'' \emph{Journal of Visual Languages \&
  Computing}, vol.~7, no.~1, pp. 33--55, 1996.

\bibitem{hetzler1998multi}
B.~Hetzler, P.~Whitney, L.~Martucci, and J.~Thomas, ``Multi-faceted insight
  through interoperable visual information analysis paradigms,'' in
  \emph{Information Visualization, 1998. Proceedings. IEEE Symposium on}.\hskip
  1em plus 0.5em minus 0.4em\relax IEEE, 1998, pp. 137--144.

\bibitem{Spence2001Information}
R.~Spence, ``Information visualization.''\hskip 1em plus 0.5em minus
  0.4em\relax New York: {A}{C}{M} {P}ress, 2001, pp. 85--88.

\bibitem{yang2002interring}
J.~Yang, M.~O. Ward, E.~Rundensteiner, \emph{et~al.}, ``Interring: An
  interactive tool for visually navigating and manipulating hierarchical
  structures,'' in \emph{Information Visualization, 2002. INFOVIS 2002. IEEE
  Symposium on}.\hskip 1em plus 0.5em minus 0.4em\relax IEEE, 2002, pp. 77--84.

\bibitem{HowellFilelight}
\BIBentryALTinterwordspacing
M.~Howell. Filelight. [Online]. Available:
  \url{http://www.methylblue.com/filelight/}
\BIBentrySTDinterwordspacing

\bibitem{carlis1998interactive}
J.~V. Carlis and J.~A. Konstan, ``Interactive visualization of serial periodic
  data,'' in \emph{Proceedings of the 11th annual ACM symposium on User
  interface software and technology}.\hskip 1em plus 0.5em minus 0.4em\relax
  ACM, 1998, pp. 29--38.

\bibitem{spoerri2004rankspiral}
A.~Spoerri, ``Rankspiral: Toward enhancing search results visualizations,'' in
  \emph{Information Visualization, 2004. INFOVIS 2004. IEEE Symposium
  on}.\hskip 1em plus 0.5em minus 0.4em\relax IEEE, 2004, pp. p18--p18.

\bibitem{hong2003zoomology}
J.~Y. Hong, J.~D’Andries, M.~Richman, and M.~Westfall, ``Zoomology: comparing
  two large hierarchical trees,'' \emph{Posters Compendium of Information
  Visualization}, pp. 120--121, 2003.

\bibitem{krzywinski2009circos}
M.~Krzywinski, J.~Schein, I.~Birol, J.~Connors, R.~Gascoyne, D.~Horsman, S.~J.
  Jones, and M.~A. Marra, ``Circos: an information aesthetic for comparative
  genomics,'' \emph{Genome research}, vol.~19, no.~9, pp. 1639--1645, 2009.

\bibitem{draper2008votes}
G.~M. Draper and R.~F. Riesenfeld, ``Who votes for what? a visual query
  language for opinion data,'' \emph{Visualization and Computer Graphics, IEEE
  Transactions on}, vol.~14, no.~6, pp. 1197--1204, 2008.

\bibitem{bostock2011d3}
M.~Bostock, V.~Ogievetsky, and J.~Heer, ``D$^3$ data-driven documents,''
  \emph{Visualization and Computer Graphics, IEEE Transactions on}, vol.~17,
  no.~12, pp. 2301--2309, 2011.

\bibitem{north2006toward}
N.~Chris, ``Toward measuring visualization insight,'' \emph{Computer Graphics
  and Applications, IEEE}, vol.~26, no.~3, pp. 6--9, 2006.

\bibitem{saraiya2004evaluation}
P.~Saraiya, C.~North, and K.~Duca, ``An evaluation of microarray visualization
  tools for biological insight,'' in \emph{Information Visualization, 2004.
  INFOVIS 2004. IEEE Symposium on}.\hskip 1em plus 0.5em minus 0.4em\relax
  IEEE, 2004, pp. 1--8.

\bibitem{roebuck2012agile}
K.~Roebuck, \emph{Agile Software Development: High-impact Strategies-What You
  Need to Know: Definitions, Adoptions, Impact, Benefits, Maturity,
  Vendors}.\hskip 1em plus 0.5em minus 0.4em\relax Brisbane, Australia: Emereo
  Publishing, 2012.

\bibitem{beck2000extreme}
K.~Beck, \emph{Extreme programming explained: embrace change}.\hskip 1em plus
  0.5em minus 0.4em\relax Boston, Massachusetts: Addison-Wesley Professional,
  2000.

\bibitem{poutanen2010validity}
O.~Poutanen, A.-M. Koivisto, S.~K{\"a}{\"a}ri{\"a}, and R.~K. Salokangas, ``The
  validity of the depression scale (deps) to assess the severity of depression
  in primary care patients,'' \emph{Family practice}, vol.~27, no.~5, pp.
  527--534, 2010.

\bibitem{MDNJS}
\BIBentryALTinterwordspacing
{The Mozilla Developer Network (MDN)}. The typeof operator - javascript
  reference. [Online]. Available:
  \url{https://developer.mozilla.org/en-US/docs/Web/JavaScript/Reference/Operators/typeof}
\BIBentrySTDinterwordspacing

\bibitem{nielsen1994usability}
J.~Nielsen, \emph{Usability Engineering}.\hskip 1em plus 0.5em minus
  0.4em\relax Amsterdam, The Netherlands: Elsevier, 1994.

\bibitem{johnson2011ehr}
C.~Johnson, D.~Johnston, P.~Crowle, \emph{et~al.}, ``Ehr usability toolkit: A
  background report on usability and electronic health records,''
  \emph{Rockville, MD: Agency for Healthcare Research and Quality}, 2011.

\bibitem{Nielsen10UsabilityHeuristics}
\BIBentryALTinterwordspacing
J.~Nielsen. (1995) 10 usability heuristics for user interface design. [Online].
  Available: \url{http://www.nngroup.com/articles/ten-usability-heuristics/}
\BIBentrySTDinterwordspacing

\bibitem{wharton1994cognitive}
C.~Wharton, J.~Rieman, C.~Lewis, and P.~Polson, ``The cognitive walkthrough
  method: A practitioner's guide,'' in \emph{Usability inspection
  methods}.\hskip 1em plus 0.5em minus 0.4em\relax John Wiley \& Sons, Inc.,
  1994, pp. 105--140.

\bibitem{peute2007significance}
L.~W. Peute and M.~W. Jaspers, ``The significance of a usability evaluation of
  an emerging laboratory order entry system,'' \emph{International journal of
  medical informatics}, vol.~76, no.~2, pp. 157--168, 2007.

\bibitem{karahoca2010information}
A.~Karahoca, E.~Bayraktar, E.~Tatoglu, and D.~Karahoca, ``Information system
  design for a hospital emergency department: A usability analysis of software
  prototypes,'' \emph{Journal of biomedical informatics}, vol.~43, no.~2, pp.
  224--232, 2010.

\bibitem{cohen2004cognitive}
T.~Cohen, D.~Kaufman, T.~White, G.~Segal, A.~B. Staub, V.~Patel, and
  M.~Finnerty, ``Cognitive evaluation of an innovative psychiatric clinical
  knowledge enhancement system,'' \emph{Medinfo}, vol.~11, no. Pt 2, pp.
  1295--9, 2004.

\bibitem{khajouei2009usability}
R.~Khajouei, D.~de~Jongh, and M.~W. Jaspers, ``Usability evaluation of a
  computerized physician order entry for medication ordering.'' in \emph{MIE},
  2009, pp. 532--536.

\bibitem{blackmon2002cognitive}
M.~H. Blackmon, P.~G. Polson, M.~Kitajima, and C.~Lewis, ``Cognitive
  walkthrough for the web,'' in \emph{Proceedings of the SIGCHI conference on
  human factors in computing systems}.\hskip 1em plus 0.5em minus 0.4em\relax
  ACM, 2002, pp. 463--470.

\bibitem{sears1997heuristic}
A.~Sears, ``Heuristic walkthroughs: Finding the problems without the noise,''
  \emph{International Journal of Human-Computer Interaction}, vol.~9, no.~3,
  pp. 213--234, 1997.

\bibitem{beuscart2007human}
M.-C. Beuscart-Z{\'e}phir, P.~Elkin, S.~Pelayo, R.~Beuscart, \emph{et~al.},
  ``The human factors engineering approach to biomedical informatics projects:
  state of the art, results, benefits and challenges,'' \emph{IMIA Yearbook},
  vol.~2, pp. 109--127, 2007.

\bibitem{newman199810}
W.~M. Newman, ``10. on simulation, measurement, and pfeeewiaie usability
  evaluation,'' \emph{Commentary on" Damaged Merchandise?}, vol.~13, p. 316,
  1998.

\bibitem{ericsson1980verbal}
K.~A. Ericsson and H.~A. Simon, ``Verbal reports as data.'' \emph{Psychological
  review}, vol.~87, no.~3, p. 215, 1980.

\bibitem{borycki2009novice}
E.~M. Borycki, L.~Lemieux-Charles, L.~Nagle, and G.~Eysenbach, ``Novice nurse
  information needs in paper and hybrid electronic-paper environments: a
  qualitative analysis.'' in \emph{MIE}, 2009, pp. 913--917.

\bibitem{currie2003clinical}
L.~M. Currie, M.~Graham, M.~Allen, S.~Bakken, V.~Patel, and J.~J. Cimino,
  ``Clinical information needs in context: an observational study of clinicians
  while using a clinical information system,'' in \emph{AMIA Annual Symposium
  proceedings}, vol. 2003.\hskip 1em plus 0.5em minus 0.4em\relax American
  Medical Informatics Association, 2003, p. 190.

\bibitem{hasman2006development}
A.~Hasman \emph{et~al.}, ``Development of methods for usability evaluations of
  ehr systems,'' in \emph{Ubiquity: Technologies for Better Health in Aging
  Societies: Proceedings of MIE2006}, vol. 124.\hskip 1em plus 0.5em minus
  0.4em\relax IOS Press, 2006, p. 341.

\bibitem{wu2008usability}
R.~C. Wu, M.~S. Orr, M.~Chignell, and S.~E. Straus, ``Usability of a mobile
  electronic medical record prototype: a verbal protocol analysis,''
  \emph{Informatics for Health and Social Care}, vol.~33, no.~2, pp. 139--149,
  2008.

\bibitem{lewis1995ibm}
J.~R. Lewis, ``Ibm computer usability satisfaction questionnaires: psychometric
  evaluation and instructions for use,'' \emph{International Journal of
  Human-Computer Interaction}, vol.~7, no.~1, pp. 57--78, 1995.

\bibitem{lewis1992psychometric}
------, ``Psychometric evaluation of the post-study system usability
  questionnaire: The pssuq,'' in \emph{Proceedings of the Human Factors and
  Ergonomics Society Annual Meeting}, vol.~36, no.~16.\hskip 1em plus 0.5em
  minus 0.4em\relax SAGE Publications, 1992, pp. 1259--1260.

\bibitem{jaspers2008pre}
M.~W. Jaspers, L.~W. Peute, A.~Lauteslager, and P.~J. Bakker, ``Pre-post
  evaluation of physicians' satisfaction with a redesigned electronic medical
  record system,'' \emph{Studies in health technology and informatics}, vol.
  136, p. 303, 2008.

\bibitem{goud2008subjective}
R.~Goud, M.~W. Jaspers, A.~Hasman, and N.~Peek, ``Subjective usability of the
  cardss guideline-based decision support system.'' \emph{Studies in health
  technology and informatics}, vol. 136, p. 193, 2008.

\bibitem{lewis1991psychometric}
J.~R. Lewis, ``Psychometric evaluation of an after-scenario questionnaire for
  computer usability studies: the asq,'' \emph{ACM SIGCHI Bulletin}, vol.~23,
  no.~1, pp. 78--81, 1991.

\bibitem{bangor2008empirical}
A.~Bangor, P.~T. Kortum, and J.~T. Miller, ``An empirical evaluation of the
  system usability scale,'' \emph{Intl. Journal of Human--Computer
  Interaction}, vol.~24, no.~6, pp. 574--594, 2008.

\bibitem{us2008physical}
U.~D. of~Health, H.~Services, \emph{et~al.}, ``Physical activity guidelines for
  americans: Be active, healthy, and happy! odphp publication no. u0036.
  october 2008,'' 2008.

\bibitem{physical2008physical}
P.~A. G.~A. Committee \emph{et~al.}, ``Physical activity guidelines advisory
  committee report, 2008,'' \emph{Washington, DC: US Department of Health and
  Human Services}, vol. 2008, p.~A5, 2008.

\bibitem{tudor2004many}
C.~Tudor-Locke and D.~R. Bassett~Jr, ``How many steps/day are enough?''
  \emph{Sports Medicine}, vol.~34, no.~1, pp. 1--8, 2004.

\bibitem{oja2013tester}
\BIBentryALTinterwordspacing
P.~Oja, A.~M{\"a}ntt{\"a}ri, T.~Pokki, K.~Kukkonen-Harjala, R.~Laukkanen, and
  J.~Malmberg, ``Tester's guide: Ukk walk test,'' 2013. [Online]. Available:
  \url{http://www.ukkinstituutti.fi/filebank/1118-UKK_walk_test_testers_guide.pdf}
\BIBentrySTDinterwordspacing

\bibitem{laukkanen1992validity}
R.~Laukkanen, P.~Oja, M.~Pasanen, and I.~Vuori, ``Validity of a two kilometre
  walking test for estimating maximal aerobic power in overweight adults.''
  \emph{International journal of obesity and related metabolic disorders:
  journal of the International Association for the Study of Obesity}, vol.~16,
  no.~4, pp. 263--268, 1992.

\bibitem{suni2009fitness}
\BIBentryALTinterwordspacing
J.~Suni, P.~Husu, and M.~Rinne, ``Fitness for health: the alpha-fit test
  battery for adults aged 18--69,'' 2009. [Online]. Available:
  \url{http://www.ukkinstituutti.fi/filebank/500-ALPHA_FIT_Testers_Manual.pdf}
\BIBentrySTDinterwordspacing

\bibitem{firstbeat2014}
\BIBentryALTinterwordspacing
{Firstbeat Technologies Ltd.} (2014) Stress and recovery analysis method based
  on 24-hour heart rate variability. white paper. [Online]. Available:
  \url{http://www.firstbeat.com/userData/firstbeat/research-publications/Stress-and-recovery_white-paper_2014.pdf}
\BIBentrySTDinterwordspacing

\bibitem{teisala2014associations}
T.~Teisala, S.~Mutikainen, A.~Tolvanen, M.~Rottensteiner, T.~Leskinen,
  J.~Kaprio, M.~Kolehmainen, H.~Rusko, and U.~M. Kujala, ``Associations of
  physical activity, fitness, and body composition with heart rate
  variability-based indicators of stress and recovery on workdays: a
  cross-sectional study,'' \emph{J Occup Med Toxicol}, vol.~34, pp. 26--40,
  2014.

\bibitem{scheier1994distinguishing}
M.~F. Scheier, C.~S. Carver, and M.~W. Bridges, ``Distinguishing optimism from
  neuroticism (and trait anxiety, self-mastery, and self-esteem): a
  reevaluation of the life orientation test.'' \emph{Journal of personality and
  social psychology}, vol.~67, no.~6, p. 1063, 1994.

\bibitem{faulkner2003beyond}
L.~Faulkner, ``Beyond the five-user assumption: Benefits of increased sample
  sizes in usability testing,'' \emph{Behavior Research Methods, Instruments,
  \& Computers}, vol.~35, no.~3, pp. 379--383, 2003.

\bibitem{berry2015usability}
D.~L. Berry, B.~Halpenny, J.~L. Bosco, J.~Bruyere, and M.~G. Sanda, ``Usability
  evaluation and adaptation of the e-health personal patient profile-prostate
  decision aid for spanish-speaking latino men,'' \emph{BMC medical informatics
  and decision making}, vol.~15, no.~1, p.~56, 2015.

\bibitem{zhang2003using}
J.~Zhang, T.~R. Johnson, V.~L. Patel, D.~L. Paige, and T.~Kubose, ``Using
  usability heuristics to evaluate patient safety of medical devices,''
  \emph{Journal of biomedical informatics}, vol.~36, no.~1, pp. 23--30, 2003.

\bibitem{shneiderman1992designing}
B.~Shneiderman, \emph{Designing the user interface: strategies for effective
  human-computer interaction}.\hskip 1em plus 0.5em minus 0.4em\relax Boston,
  Massachusetts: Addison-Wesley Reading, MA, 1992, vol.~3.

\bibitem{molich1990improving}
R.~Molich and J.~Nielsen, ``Improving a human-computer dialogue,''
  \emph{Communications of the ACM}, vol.~33, no.~3, pp. 338--348, 1990.

\bibitem{Tognazzini2014principles}
\BIBentryALTinterwordspacing
B.~Tognazzini. (2014) First principles of information design. [Online].
  Available: \url{http://www.asktog.com/basics/firstPrinciples.html}
\BIBentrySTDinterwordspacing

\bibitem{tenk2009}
\BIBentryALTinterwordspacing
\emph{Ethical principles of research in the humanities and social and
  behavioural sciences and proposals for ethical review}, Finnish Advisory
  Board on Research Integrity, Helsinki, 2009. [Online]. Available:
  \url{http://www.tenk.fi/en/ethical-review-human-sciences}
\BIBentrySTDinterwordspacing

\bibitem{lewis2002psychometric}
J.~R. Lewis, ``Psychometric evaluation of the pssuq using data from five years
  of usability studies,'' \emph{International Journal of Human-Computer
  Interaction}, vol.~14, no. 3-4, pp. 463--488, 2002.

\bibitem{aight}
\BIBentryALTinterwordspacing
S.~Allen. Aight -- {JavaScript} shims and shams for making {IE8-9} behave
  reasonably. [Online]. Available: \url{https://github.com/shawnbot/aight}
\BIBentrySTDinterwordspacing

\bibitem{rind2011interactive}
A.~Rind, T.~D. Wang, A.~Wolfgang, S.~Miksch, K.~Wongsuphasawat, C.~Plaisant,
  and B.~Shneiderman, ``Interactive information visualization to explore and
  query electronic health records,'' \emph{Foundations and Trends in
  Human-Computer Interaction}, vol.~5, no.~3, pp. 207--298, 2011.

\end{thebibliography}

\end{document}